\documentclass[aps, prb, twocolumn, amssymb, amsmath, showpacs, superscriptaddress]{revtex4-1}

\usepackage{bm}
\usepackage{times}
\usepackage{graphicx}
\usepackage{color}
\usepackage{dcolumn}
\usepackage[colorlinks=true, letterpaper=true, pdfstartview=FitV, linkcolor=blue, citecolor=blue, urlcolor=blue]{hyperref}
\usepackage{appendix}
\usepackage[normalem]{ulem}

\begin{document}
\title{Gauge invariant quantum transport theory for non-Hermitian systems}
\author{Miaomiao Wei}
\affiliation{College of Physics and Optoelectronic Engineering, Shenzhen University, Guangdong, P. R. China}
\author{Bin Wang}
\email[]{bwang@szu.edu.cn}
\affiliation{College of Physics and Optoelectronic Engineering, Shenzhen University, Guangdong, P. R. China}
\author{Jian Wang}
\email[]{jianwang@hku.hk}
\affiliation{College of Physics and Optoelectronic Engineering, Shenzhen University, Guangdong, P. R. China}
\affiliation{Department of Physics, University of Hong Kong, Pokfulam Road, Hong Kong, P. R. China}
\begin{abstract}
Gauge invariance is a fundamental principle that must be preserved in quantum transport. However, when a complex potential is incorporated into the Hamiltonian, we find that the current described by the well-established Landauer-B$\ddot{u}$ttiker formula no longer satisfies gauge invariance. Using the non-equilibrium Green's function (NEGF) method, we derive a current expression for a multi-probe system that includes a complex potential in the scattering region. We observe that an additional current term arises compared to the Landauer-B$\ddot{u}$ttiker formula, which leads to a violation of gauge invariance. To address this, we propose two phenomenological methods for redistributing the conductance to restore gauge invariance in non-Hermitian systems. These methods are applied to various trivial and nontrivial non-Hermitian quantum states, confirming the necessity of gauge-invariant treatments in non-Hermitian systems.
\end{abstract}
	\maketitle
\section{Introduction}
In a closed quantum system, the Hamiltonian is typically Hermitian, leading to unitary evolution. However, in an open quantum system that interacts with the environment, gain and loss emerge, necessitating a more comprehensive description beyond purely Hermitian processes\cite{PRL.123.056801,PRL.123.056802,RMP.93.015005,AdP.69.249,PRL.121.086803,Lindblad, book1}.  In such cases, the system's evolution is governed by both continuous dynamics, captured by a non-Hermitian Hamiltonian, and discrete quantum jumps, with the system's density matrix evolving according to the Lindblad master equation\cite{Lindblad, book1}.
On short timescales, where quantum jumps have not yet occurred, the system's evolution can be modeled using a non-Hermitian Hamiltonian\cite{example1}. The non-Hermitian Hamiltonian describes the continuous, deterministic evolution of the system, encompassing phenomena such as the exponential decay of unstable states and the initial dynamics in systems involving gain or loss mechanisms. In particular, for a non-Hermitian system possessing Parity-Time  ($\mathcal{PT}$) symmetry, the eigenvalues remain real, supporting stable and predictable dynamics within specific phases. Unique phenomena associated with  $\mathcal{PT}$-symmetry, such as the non-Hermitian skin effect\cite{PRB.99.201103,NP.16.761,NC.12.6297,PRL.113.103903,PRL.118.116801} and robust edge states\cite{PRL.121.026808,PRL.362.325,PRR.6.013213,PRB.97.121401}, underscore the critical role of  $\mathcal{PT}$-symmetry in non-Hermitian physics.

Quantum transport in non-Hermitian systems is a rapidly advancing field that bridges theoretical predictions with experimental realizations \cite{NJP.10.113019, RPP.79.096001, PRL.122.050501, PRR.5.033107, PRL.132.136301, AOP.15.4422, NP.11.752, PRL.129.070401, PRL.124.250402, Adv.78905}. By introducing gain and loss mechanisms, non-Hermitian systems exhibit a range of  novel transport phenomena, such as unidirectional transport, exceptional points, and non-Hermitian topological phases \cite{PRL.113.103903, PRL.118.116801, PRL.121.026808, PRL.362.325, PRR.6.013213}. Experimental platforms, including optical systems \cite{AOP.15.4422, NP.11.752}, cold atoms \cite{PRL.129.070401, PRL.124.250402} and condensed matter \cite{Adv.78905},  provide various means to explore these effects, promoting the understanding of quantum transport in non-Hermitian systems. However, a fundamental theoretical issue in quantum transport within non-Hermitian systems remains unaddressed --- gauge invariance. Gauge invariance dictates that the current in each lead should remain unchanged when the bias at each lead is shifted by a constant value. Different from current conservation, which can be violated due to the presence of complex potentials, gauge invariance must be satisfied to ensure the consistency and validity of physical laws, irrespective of the chosen gauge for electromagnetic fields.

In the weakly nonlinear regime, the current can be expressed as\cite{PRL.81.2763}
\begin{equation}
I_\alpha=\sum_\beta G_{\alpha\beta}v_\beta+\sum_{\beta\gamma}G_{\alpha\beta\gamma}v_\beta v_\gamma + ..., \nonumber
\label{curr1}
\end{equation}
where $G_{\alpha\beta}$ represents the linear conductance, $G_{\alpha\beta\gamma}$ denotes the second order nonlinear conductance, and $v_\beta$ is the voltage in lead $\beta$.
Gauge invariance refers to the constraint $\sum_\beta G_{\alpha\beta} = 0$ in the linear case \cite{PRL.81.2763,footnote}, and \(\sum_{\beta} (G_{\alpha\beta\gamma} + G_{\alpha\gamma\beta}) = 0\) in the second-order nonlinear case\cite{PRL.81.2763}. In Hermitian systems, the linear conductance constraint is naturally fulfilled, and the nonlinear constraint is satisfied by incorporating the Coulomb potential into the system\cite{JPCM.5.9361, JAP.86.5094,jianpaper}. However, in non-Hermitian systems, we find that even the linear conductance constraint can be violated due to the non-unitary nature of the evolution operator, leading to a failure of gauge variance. Nonetheless, the well-established Landauer-B$\ddot{u}$ttiker formula, originally derived for Hermitian systems, has been directly applied to non-Hermitian systems in some studies\cite{PRA.95.062123, PRA.99.032119, PRB.105.125103, PRB.106.045142}. Therefore, it is essential to develop current and conductance formulas applicable to non-Hermitian systems while preserving gauge invariance.

Dephasing in quantum transport for Hermitian systems, which refers to the loss of quantum coherence, has been extensively studied and is somewhat analogous to non-Hermitian electron dynamics \cite{Price, Datta2,jianpaper2}. Therefore, it is necessary to review the methods applied to address dephasing in quantum transport. One conventional approach is the introduction of a virtual probe into the system \cite{Buttiker1, Datta1}. In this method, electrons are allowed to exchange between the scattering region and the virtual probe, while ensuring that the net current in the virtual probe remains zero. During this process, the electron's phase is lost due to thermalization in the virtual probe \cite{M-Wei1}. Another effective method for describing dephasing in quantum transport is the current partition method, where the total current is divided into elastic and inelastic components \cite{PRL.82.398}. The elastic current in each lead is explicitly expressed by the Landauer-B$\ddot{u}$ttiker formula, while only the total inelastic current is provided. To ensure current conservation and gauge invariance, the total inelastic current must be appropriately partitioned among the leads. Currently, only a phenomenological theory exists for this partition method \cite{PRL.82.398}.
	
In this work, we derive the current expression using the non-equilibrium Green's function (NEGF) method in a non-Hermitian system, where a non-Hermitian quantum dot is connected to multiple Hermitian leads. The non-Hermitian term is treated using the virtual probe method,  analogous to how dephasing is handled\cite{Buttiker1, Datta1}. Beyond the well-known Landauer-B$\ddot{u}$ttiker formula, the presence of non-Hermitian terms introduces additional components to the current.
As a result, gauge invariance may not be satisfied, depending on $G_{\alpha F}$, which represents the conductance between the real lead $\alpha$ and the virtual probe $F$.
If $G_{\alpha F} = 0$, the current reduces to the Landauer-B$\ddot{u}$ttiker formula, naturally satisfying both gauge invariance and current conservation. However, if $G_{\alpha F} \neq 0$, the virtual probe method or the current partition method should be employed to ensure gauge invariance. The derived formulas for current and conductance were applied to various non-Hermitian systems, and numerical verification confirms the necessity of current partition for maintaining gauge invariance.
		
This paper is organized as follows. In Section II, we derive the current expression for a multi-probe non-Hermitian system based on the NEGF method and present the virtual probe method and current partition method to ensure the gauge invariance of conductance. In Section III, the proposed formulas are utilized to calculate the conductances of several non-Hermitian systems. Finally, in Section IV, we present a summary of the findings.

\section{Theoretical Formalism}
We study the quantum transport phenomena within a non-Hermitian system where a non-Hermitian quantum dot is connected by multiple Hermitian leads. The Hamiltonian of the system is given by
\begin{eqnarray}
H = \sum_\alpha H_\alpha + H_{dot} + \sum_\alpha H_{\alpha T},  \nonumber
\end{eqnarray}
where
\begin{eqnarray}
H_\alpha &=& \sum_{k}\epsilon_{k\alpha}c_{k\alpha}^\dagger c_{k\alpha}   \nonumber \\
H_{dot} &=& \sum_n \epsilon_n d_n^\dagger d_n, \nonumber \\
H_{\alpha T} &=& \sum_{k n} t_{k\alpha n} c_{k\alpha}^\dagger d_n + h.c.. \nonumber
\label{Hsys}
\end{eqnarray}
Here, $H_\alpha$ represents the Hamiltonian of lead $\alpha$, where $c_{k\alpha}^\dagger (c_{k\alpha})$ is the creation (annihilation) operator for the $k$-state in lead $\alpha$, and $\epsilon_{k\alpha}$ is the corresponding energy spectrum. $H_{dot}$ denotes the Hamiltonian of the quantum dot with $ d_n^\dagger (d_n) $ being the creation (annihilation) operator for the $n$-th energy level $\epsilon_n$. If $\epsilon_n$ is real, the system is Hermitian. However, if $\epsilon_n$ is complex, the system becomes non-Hermitian. The Hamiltonian matrix for the non-Hermitian quantum dot can be written as
\begin{equation}
H_{dot} = H_0 -i\Gamma_F/2, \nonumber \\
\end{equation}
where $H_0$ is a Hermitian matrix, and $\Gamma_F$ is a real matrix describing the strength of non-Hermitian term. 
There are various mechanisms that contribute to non-Hermiticity, such as energy gain and loss\cite{prl.130.163001, prl.129.093001}, non-reciprocal coupling\cite{np.20.395, nc.15.1798}, unconventional Goos-H$\ddot{a}$nchen effects\cite{PRL.129.086601}, and dephasing\cite{Price, Datta2,jianpaper2}. However, in this paper, we focus only on the theoretical investigation of the impact of the non-Hermitian term $i\Gamma_F/2$ on the current passing through the system, without delving into the physical origins of the non-Hermiticity.

\subsection{current formula}
In general, a microscopic expression for current in a non-Hermitian system should be derived from the Lindblad master equation\cite{Lindblad, book1}. Although several theories have been developed to describe quantum transport in non-Hermitian systems \cite{PRL.81.2763, PRA.95.062123, PRA.99.032119, PRB.105.125103, PRB.106.045142,arxiv.08973}, they are primarily suitable for periodic systems and are inadequate for open systems with multiple leads. Due to the complexity of the theoretical derivation, a microscopic current expression for non-Hermitian systems with multiple leads has remained undeveloped until recently. As a result, the Landauer-B$\ddot{u}$ttiker formula, originally developed for Hermitian systems, has often been directly applied to non-Hermitian systems \cite{PRA.95.062123, PRA.99.032119, PRB.105.125103, PRB.106.045142}. Recently, a microscopic current expression was derived from the motion equation for a non-Hermitian quantum dot system connected to two reservoirs\cite{PRL.132.136301}. This theory reveals that the current in a non-Hermitian system consists two components: an elastic scattering term, as described by the Landauer-B$\ddot{u}$ttiker formula, and an inelastic scattering term that depends on the non-Hermitian strength. However, this theory does not satisfy gauge invariance, as we will discuss later.

In the following, we derive the current from the general expression given by Eq. (\ref{current}), which is based on the motion equation for a Hermitian system. For Hermitian systems, the Landauer-B$\ddot{u}$ttiker formula is naturally derived from Eq. (\ref{current}). However, for non-Hermitian systems, Eq. (\ref{current}) leads to a different current expression, which we will demonstrate is structurally equivalent to the microscopic theory presented in Ref.[\onlinecite{PRL.132.136301}] under certain approximations. Our starting point is the expression for the current\cite{PRL.68.2512}
\begin{equation}
I_\alpha = -2i\int_E {\bf Tr}[\textrm{Im}(\Sigma_\alpha^r) (G^r-G^a) f_\alpha + \textrm{Im}(\Sigma_\alpha^r) G^<],
\label{current}
\end{equation}
where $\int_E  = \int dE/2\pi$ and $f_\alpha = f(E - v_\alpha)$ represents the Fermi distribution function with $v_\alpha$ being the bias of lead $\alpha$. $G^r$ is the retarded Green's function defined as
\begin{eqnarray}
G^r = \frac{1}{E- H_0 - \sum_\alpha \Sigma^r_\alpha + i\Gamma_F/2}, \label{gr0}
\end{eqnarray}
where $\Sigma^r_{\alpha}$ represents the self-energy of lead $\alpha$, and $G^a=[G^r]^\dagger$.
Although Eq. (\ref{current}) was originally derived for a Hermitian Hamiltonian, it has been successfully applied to describe dephasing in quantum transport, where a non-Hermitian term $i\Gamma_F/2$ is introduced into $G^r$, representing dephasing strength, similar to Eq.(\ref{gr0}).
	
In Eq.(\ref{current}), the lesser Green's function $G^<$ is determined by the Keldysh equation as \cite{Jauho}
\begin{eqnarray}
G^<= G^r [(G^r_0)^{-1} G^<_0 (G^a_0)^{-1} + \Sigma^< ]G^a,
\label{lesser}
\end{eqnarray}
where $G^r_0 = 1/(E- H_0 + i\Gamma_F/2)$ and $G^<_0 = -f (G^r_0 - G^a_0)$ represent the Green's functions of the isolated system,  and $\Sigma^<=i\sum_\alpha \Gamma_\alpha f_\alpha$. It is straightforward to show that Eq.(\ref{lesser}) is equivalent to
\begin{eqnarray}
G^<= G^r [i f \Gamma_F + \Sigma^< ]G^a = G^r {\tilde\Sigma}^< G^a.
\label{lesser1}
\end{eqnarray}
The non-Hermitian term $\Gamma_F$ serves as an extra ``reservoir" with zero chemical potential, functioning as a line-width function for a fictitious lead.  By substituting Eqs.(\ref{gr0}) and (\ref{lesser1}) into Eq.(\ref{current}), we obtain
\begin{eqnarray}
I_\alpha = \int_E \sum_\beta && {\bf Tr}[\Gamma_\alpha G^r  \Gamma_\beta G^a](f_\alpha -f_\beta)+  \nonumber \\
			&& {\bf Tr}[\Gamma_\alpha G^r  \Gamma_F G^a](f_\alpha -f_F)
\label{current1},
\end{eqnarray}
where $f_F$ is the Fermi distribution function of the fictitious lead. The first term on the right-hand side of Eq. (\ref{current1}) corresponds to the well-established Landauer-B$\ddot{u}$ttiker formula, providing a familiar framework for interpreting current behavior. The influence of non-Hermitian terms is embedded within the Green's function, revealing a complex interplay between single-particle gain or loss mechanisms and elastic scattering processes. The second term represents the inelastic current due to the coupling between the quantum dot and its environment. Notably, this term indicates the emergence of an additional current, with its magnitude depending on the non-Hermitian strength $\Gamma_F$, even in the absence of a bias voltage between real leads.

We now show that Eq.(\ref{current1}) shares essential features with the microscopic expression derived in Ref.[\onlinecite{PRL.132.136301}], where the current consists of both elastic and inelastic components, aligning with Eq. (\ref{current1}). The elastic component corresponds to the Landauer-B$\ddot{u}$ttiker formula, while the inelastic component is formulated as\cite{PRL.132.136301}
\begin{eqnarray}
I^{in}_\alpha = -2\int _E \gamma {\bf Tr}[\Gamma_\alpha G^r  O(\mathcal{D}-f_\alpha)O G^a],
\label{current2}
\end{eqnarray}
where $\gamma$ represents the monitoring strength, and $O = \sum_{ij}d^\dagger_i O_{ij} d_j$ denotes the observable single-particle quantity. Here,  $d^\dagger_i$($d_i$) is the creation (annihilation) operator for the $i$-th quantum dot state, and $O_{ij}$ describes the transition probability from $j$-state to $i$-state.
According to the definition $G^r=1/(E-H-\sum_\alpha \Sigma_\alpha^r+i\gamma O^2)$, it is evident that $\gamma O^2$ is equivalent to $\Gamma_F/2$. The dynamic correlation matrix $\mathcal{D}$ can be solved self-consistently as follows\cite{PRL.132.136301}
\begin{equation}
\mathcal{D}=\int \frac{d\omega}{\pi} G^r\big[\sum_\alpha f_\alpha \Gamma_\alpha + \gamma O\mathcal{D}O\big]G^a.
\label{DD}
\end{equation}
Notably, $\gamma O\mathcal{D}O$ shares a similar role to $f_\alpha\Gamma_\alpha$, representing a lesser self-energy stemming from the non-Hermition term. Given the equivalence between $\gamma O^2$ and $\Gamma_F/2$, $\mathcal{D}$ can be interpreted as a Fermi distribution function of the non-Hermition reservoir.
By applying a unitary transformation to $\mathcal{D}$, it transforms into a diagonal matrix, i.e., $U^\dagger\mathcal{D}U=[D_{ij}\delta_{ij}]$. Assuming $D_i=f_F$, the matrix $[D_{ij}\delta_{ij}]$ simplifies to $f_FI$, where $I$ is the identity matrix. Consequently, Eq.(\ref{current2}) reduces to
\begin{eqnarray}
I^{in}_\alpha
&=& -2\int_E \gamma {\bf Tr}[\Gamma_\alpha G^r  O(f_FI-f_\alpha)O G^a]    \nonumber \\
&=& -\int_E  {\bf Tr}[\Gamma_\alpha G^r  \Gamma_F G^a (f_F-f_\alpha)].
\label{current3}
\end{eqnarray}
This result coincides with the second term on the right-hand side of Eq. (\ref{current1}) .

	
Next, we discuss the gauge invariance of Eq.(\ref{current1}). In the linear regime (small bias $v_\alpha$) and at zero temperature ($T=0$), Eq.(\ref{current1}) simplifies to
\begin{eqnarray}
I_\alpha &=& \sum_{ \beta} {\bf Tr}[\Gamma_\alpha G^r (\Gamma_\beta - {\tilde \Gamma}\delta_{\alpha\beta}) G^a]v_{\beta} = \sum_{ \beta} G_{\alpha\beta}v_{\beta}
\label{current1-2}
\end{eqnarray}
with ${\tilde \Gamma} = \sum_\alpha \Gamma_\alpha + \Gamma_F$. It is straightforward to show that
\begin{eqnarray}
\sum_\beta G_{\alpha\beta}
= -{\bf Tr}[\Gamma_\alpha G^r \Gamma_F G^a] = -G_{\alpha F}. \label{Gab2}
\end{eqnarray}
When $\Gamma_F=0$, Eq.(\ref{Gab2}) reduces to $\sum_\beta G_{\alpha\beta}=0$, meaning that gauge invariance is naturally satisfied for Hermitian systems. However, for non-Hermitian systems where $\Gamma_F\neq0$, $G_{\alpha F}$ is usually non-zero, and gauge invariance is violated.
The violation of gauge invariance also occurs in the microscopic expression derived in Ref.[\onlinecite{PRL.132.136301}]. It can be easily shown that $\sum_\beta G_{\alpha\beta} = -2{\bf Tr}[\Gamma_\alpha G^r \gamma O^2 G^a]$. When $\gamma \neq 0$, $\sum_\beta G_{\alpha\beta}$ is generally nonzero, resulting in a violation of gauge invariance.
In such cases, either the virtual probe method or the current partition method should be employed to correct the current and conductance in Eq.(\ref{current1-2}). However, in some special cases, as will be discussed in subsection C, $G_{\alpha F}$ remains zero even when $\Gamma_F$ is nonzero, meaning that gauge invariance is naturally preserved, and Eq.(\ref{current1}) reduces to the familiar Landauer-B$\ddot{u}$ttiker formula.
	
\subsection{gauge invariant methods for conductance}
In the following, we present the N-virtual probe method for ensuring gauge invariance in quantum transport within non-Hermitian systems. This approach is similar to the method used for dealing with dephasing in Hermitian systems, but with two key differences\cite{Buttiker1,Datta1,M-Wei1}. First, in Hermitian system, the exact form of the dephasing mechanism is often known,  and dephasing is therefore not included in the original Hamiltonian, but is instead phenomenologically represented by virtual probes. In contrast, in non-Hermitian systems, the exact form of dissipation is known and acts like an additional probe or inelastic transmission channel with zero bias. Second, $\Gamma_F$ is diagonal when describing dephasing in Hermitian systems, whereas it may be a full matrix for non-Hermitian systems. Assuming there are $N$ virtual probes and $n$ real leads, the current in each real lead $I_{\alpha}$ and virtual probe $I_i$ is given by\cite{PRL.81.2763}
\begin{eqnarray}
I_\alpha &=& \sum_{\beta=1:n} G_{\alpha \beta} v_\beta + \sum_ {j=1:N} G_{\alpha j} v_j  \nonumber\\
... \nonumber \\
I_i &=& \sum_{\beta=1:n} G_{i\beta} v_\beta + \sum_{j=1:N} G_{ij} v_j  \nonumber\\
... ,
\label{current5}
\end{eqnarray}
where $v_j$ is the voltage of the $j$-th virtual probe. The conductance is defined similar to Eq.(\ref{current1-2}) as
\begin{eqnarray}
G_{ab} = {\rm Tr}[\Gamma_a G^r (\Gamma_b - {\tilde \Gamma} \delta_{ab}) G^a ] \label{GFi}
\end{eqnarray}
where $a(b)$ includes both real leads $\alpha(\beta)$ and virtual probes $i(j)$. $\Gamma_i $ represents the contribution from the $i$-th virtual probe, which is an $N\times N$ matrix containing only the $i$-th row of $\Gamma_F$. Since the current in the virtual probes is not physical, $I_i$ should be set to zero by adjusting $v_j$. Expressing $v_i=\sum_\beta c_\beta^i v_\beta$ and substituting it into Eq.(\ref{current5}), we get
\begin{eqnarray}
I_i = \sum_\beta (G_{i\beta} + \sum_j G_{ij} c_\beta^j) v_\beta \equiv 0,  \nonumber
\end{eqnarray}
which determines the value of $c^j_\beta$.
Because $v_\beta$ is an arbitrary value, we have
\begin{eqnarray}
G_{i\beta} + \sum_j G_{ij} c_\beta^j = 0,  \nonumber
\end{eqnarray}
and thus
\begin{eqnarray}
c_\beta^j = -[(G_{FF})^{-1}G_{F\beta}]_{j}, \nonumber
\end{eqnarray}
where $G_{FF}$ is an $N \times N$ matrix with elements $G_{ij}$, and $G_{F\beta}$ is an $N \times 1$ column vector with elements $G_{i\beta}$. Substituting $c_\beta$ into $I_\alpha$ from Eq.(\ref{current5}), the current in the real lead $\alpha$ is given by
\begin{eqnarray}
I_\alpha = \sum_\beta [G_{\alpha \beta} - \sum_j G_{\alpha j} (G_{FF}^{-1}G_{F\beta})_{j}] v_\beta. \nonumber
\end{eqnarray}
This expression includes both elastic and inelastic currents.
Finally, we have
\begin{eqnarray}
{\tilde G_{\alpha\beta}}&=& G_{\alpha \beta} - G_{\alpha F} G_{FF}^{-1}G_{F\beta},
\label{cond1}
\end{eqnarray}
where $G_{\alpha F}$ is a $1 \times N$ row vector with elements $G_{\alpha i}$. According to the definition in Eq.(\ref{GFi}), it is straightforward to show that $\sum_{\beta} G_{\alpha\beta} = -\sum_j G_{\alpha j}$ and $\sum_{\beta} G_{i\beta} = -\sum_j G_{ij}$. Combining this with Eq.(\ref{Gab2}), we can prove that the modified conductance in Eq.(\ref{cond1}) satisfies the gauge invariance condition $\sum_\beta \tilde G_{\alpha\beta} = 0$, thus ensuring current conservation $\sum_\alpha \tilde G_{\alpha\beta} = 0$. Similar method has been successfully applied to quantum transport problems involving dephasing\cite{Y-Xing, H-Jiang, M-Wei1}, demonstrating its effectiveness in preserving gauge invariance.
	
Current partition is another effective method for obtaining correct conductance by assigning the inelastic current to each physical channel using the gauge invariant condition, which was first proposed to allocate the displacement current $I^d$ to contributions from individual probes in a dynamic transport system\cite{PRL.82.398}.  After applying current partition, the dynamic conductance satisfies both current conservation and gauge invariance. This formalism presents significant advancements: (i) it imposes no frequency limit, and (ii) it allows for detailed treatments of interactions in the mesoscopic region. For the non-Hermitian system,  assuming the assigned current can be written as ${\tilde I_{\alpha}}=I_{\alpha }+\sum_i c_\alpha^i I_i^{in}$, where ${I_{\alpha}}$ is the elastic current given by the Landauer-B$\ddot{u}$ttiker formula, $I_i^{in} = -\sum_\alpha I^i_\alpha = -\sum_\alpha G_{i\alpha}v_\alpha$ is inelastic current, and $c_\alpha$ is a $1\times N$ vector with $\sum_\alpha c_\alpha = 1$, it is easy to obtain that in the linear regime
\begin{eqnarray}
{\tilde G_{\alpha\beta}}=G_{\alpha\beta }+\sum_{i=1:N} c_\alpha^i G_{i\beta}.   \label{Glr}
\end{eqnarray}
Enforcing the gauge invariant condition $\sum_\beta {\tilde G}_{\alpha\beta} = 0$, we have $ \sum_i G_{\alpha i} + \sum_{ij}c_\alpha^i G_{ij} = 0$. Thus, we have $c_\alpha =-G_{\alpha F}G_{FF}^{-1}$.
Finally, substituting the expression for $c_\alpha$ into Eq.(\ref{Glr}), we arrive at
\begin{eqnarray}
{\tilde G}_{\alpha\beta} = G_{\alpha\beta} - G_{\alpha F}G_{FF}^{-1}G_{F\beta},
\label{cond2}
\end{eqnarray}
which is the same result obtained using the virtual probe method.

\subsection{Applicable range of gauge invariant conductance}
Subsequently, we explore the range of applicability of Eq. (\ref{cond2}). Using the Dyson equation, the system Green's function $G^r$ defined in Eq.(\ref{gr0}) can be formulated as $G^r=G^r_0+G^r_0\Sigma^r G^r$, where $G^r_0$ is the retarded Green's function of the non-Hermitian isolated quantum dot, and $\Sigma^r=\sum_\alpha \Sigma_\alpha^r$ denotes the self-energy including the contributions from all real leads. $G_0^r$ is defined as $G^r_0 = 1/(E-H_0+i\Gamma_F/2)$, which can be expressed in terms of the eigenvectors of the Hamiltonian as follows\cite{Jauho}
\begin{equation}
G^r_0 = \sum_n \frac{|\psi_n\rangle \langle\phi_n|}{E-\epsilon_n},
\label{gr00}
\end{equation}
where $|\psi_n\rangle$, $|\phi_n\rangle$ and $\epsilon_n$ are solutions of the following equations
\begin{align*}
[H_0-i\Gamma_F/2]|\psi_n\rangle   = \epsilon_n |\psi_n\rangle   \\
[H_0+i\Gamma_F/2]|\phi_n\rangle   = \epsilon^*_n |\phi_n\rangle.
\end{align*}
Typically, $\epsilon_n$ is complex, but in special cases, it becomes real. When $\epsilon_n$ is real, $|\psi_n\rangle $ and $|\phi_n\rangle$ are identical, and $G^r_0$ takes the same form as the Green's function for a Hermitian isolated quantum dot. In other words, the influence of non-Hermitian term $i\Gamma_F/2$ in $G_0^r$ and $G^r$ vanishes naturally, allowing the established Landauer-B$\ddot{u}$ttiker formula to emerge from Eq.(\ref{current}). Consequently, $G_{\alpha F}$ equals zero, and the gauge invariance of $\sum_\beta G_{\alpha\beta}$ is inherently preserved. Obviously, the reality or complexity of $\epsilon_n$ determines the applicability of Eq.(\ref{cond2}).

As pointed out in the literatures, when a non-Hermitian Hamiltonian belongs to the class of pseudo-Hermitian systems\cite{Mostafazadeh,Batal,Ghatak} or exhibits $\mathcal{PT}$-symmetry\cite{PRB.99.201103,NP.16.761, PRL.113.103903,PRL.118.116801,NC.12.6297,PRL.121.026808,PRB.97.121401,PRL.362.325,PRR.6.013213}, its eigenvalues can be real under certain conditions. For a pseudo-Hermitian Hamiltonian, the relation $\eta^{-1}(H_0-i\Gamma/2)^\dagger\eta = H_0-i\Gamma/2$ holds, where $\eta$ is the the metric operator with the canonical form $\eta=\sum_n|\phi_n \rangle \langle \phi_n |$ and $\eta^{-1}=\sum_n|\psi_n \rangle \langle \psi_n |$\cite{Ghatak}. If $\eta$ simplifies to the identity, the pseudo-Hermitian Hamiltonian reduces to a Hermitian one. Additionally, when $|\psi_n\rangle$ and $|\phi_n\rangle$ are parallel (i.e., $\eta^{-1}\phi_n=c_n\psi_n$ with $c_n$ a complex number), the eigenvalues of $H_0-i\Gamma/2$ become real. A $\mathcal{PT}$-symmetric non-Hermitian Hamiltonian is a subset of pseudo-Hermitian Hamiltonian. For a $\mathcal{PT}$-invariant Hamiltonian $H_0-i\Gamma/2$ obeying the condition $\mathcal{PT}(H_0 - i\Gamma/2)(\mathcal{PT})^{-1} = H_0 - i\Gamma/2$, it can possess real eigenvalues, provided that the $(\mathcal{PT})^2 = +1$ symmetry is present\cite{Ghatak}.

In the following, we numerically examine the validity of the gauge invariant conductance, as described by Eq.(\ref{cond2}), in different systems with various non-Hermitian terms.

\section{Numerical results}
In this section, we introduce various non-Hermitian perturbations into different systems to examine the necessity of gauge invariance formula in Eq.(\ref{cond1}). In each system, a complex potential is incorporated into the quantum dot, whereas the Hamiltonians of both leads remain Hermitian.
		
\begin{figure}[t]
\includegraphics[width=8cm]{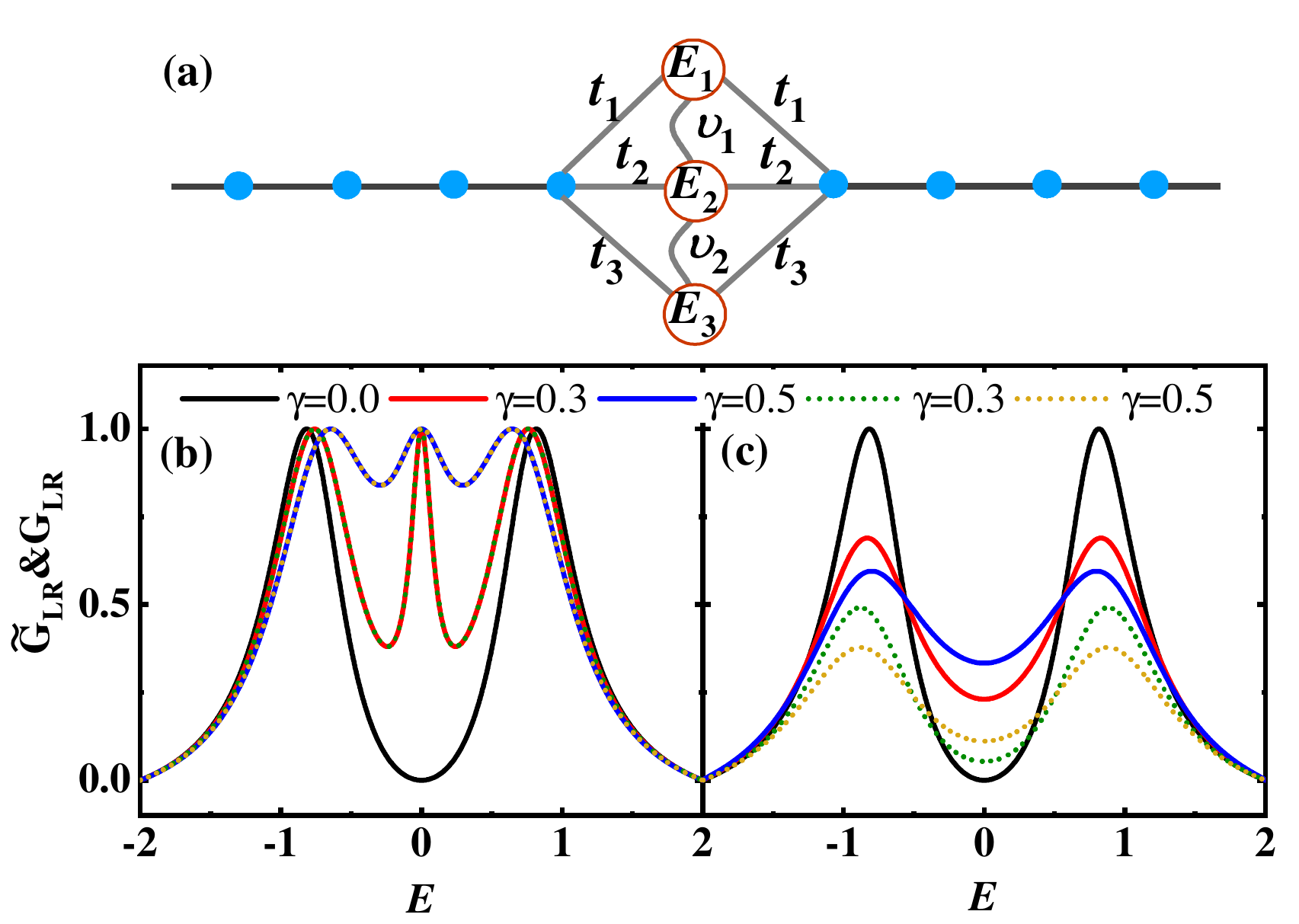}
\centering
\caption{(Color online) (a) Schematic structure of  a triple-quantum-dot system. $t_1$, $t_2$ and $t_3$ represent the interaction between the leads and the three quantum dots, while $v_1$ and $v_2$ denote the coupling strength between adjacent quantum dots.  (b) and (c) show $G_{LR}$ and $\tilde G_{LR}$ versus energy $E$ with different non-Hermitian strength $\gamma$. In (b), $E_1=E_0+i\gamma$ and $E_3=E_0-i\gamma$; In (c), $E_1=E_3=E_0+i\gamma$. The solid curves describe $\tilde G_{LR}$, while the dashed curves indicate $G_{LR}$. In both cases, we set $t_1=t_3=0$, $t_2=0.5$, $v_1=v_2=0.5$ and $E_0=0$.}
\label{fig1}
\end{figure}
		
The first system is a triple-quantum-dot structure, depicted in Fig.1(a), where three quantum dots (QD) are coupled to two metallic leads. The Hamiltonian of three QDs is defined as\cite{PRA.95.062123}
\begin{eqnarray}
H_{dot} &=& \sum_{n=1,2,3}\epsilon_n d_n^\dagger d_n + \sum_{n=1,2}v_n d_{n+1}^\dagger d_{n} + h.c.,
\end{eqnarray}
where $\epsilon_n$ represents the energy of the $n$-th QD, and $v_n$ denotes the coupling strength between adjacent QDs.  When all $\epsilon_n$ values are real, the Hamiltonian is Hermitian. However, when $\epsilon_n$ becomes complex, the Hamiltonian becomes non-Hermitian. Initially, we set $E_1=E_2=E_3=E_0$, and the system is $\mathcal{PT}$-symmetric. Then, two types of complex potential are incorporated to $E_1$ and $E_3$. In the first case, by changing $E_1=E_0+i\gamma$ and $E_3=E_0-i\gamma$, the $\mathcal{PT}$ symmetry of $H_{dot}$ is maintained\cite{PRA.95.062123}. In the second case, by setting $E_1=E_3=E_0-i\gamma$, the $\mathcal{PT}$ symmetry of $H_{dot}$ is broken. Here, $\gamma$ is a real number, indicating the strength of the non-Hermitian term.
		
The conductances $G_{LR}$ and $\tilde G_{LR}$ with different $\gamma$ are presented in Fig.1(b) and 1(c), respectively, for the first and second non-Hermitian cases. In the first case with a $\mathcal{PT}$-symmetric Hamiltonian, where $E_1=-E_3=i\gamma$, $\tilde G_{LR}$ is exactly equal to $G_{LR}$, regardless of the strength of the complex potential. The inelastic scattering current described by Eq.(\ref{current1}) vanishes, and the current calculated using the Laudauer-B$\ddot{u}$ttiker formula automatically satisfies gauge invariance. In the second case, with $\mathcal{PT}$-symmetry-broken Hamiltonian where $E_1=E_3=-i\gamma$, $\tilde G_{LR}$ varies with the magnitude of $\gamma$. Particularly, $\tilde G_{LR}$ is significantly different from $G_{LR}$, highlighting the necessity of current partition to ensure the gauge invariance.
		
\begin{figure}[t]
\includegraphics[width=8cm]{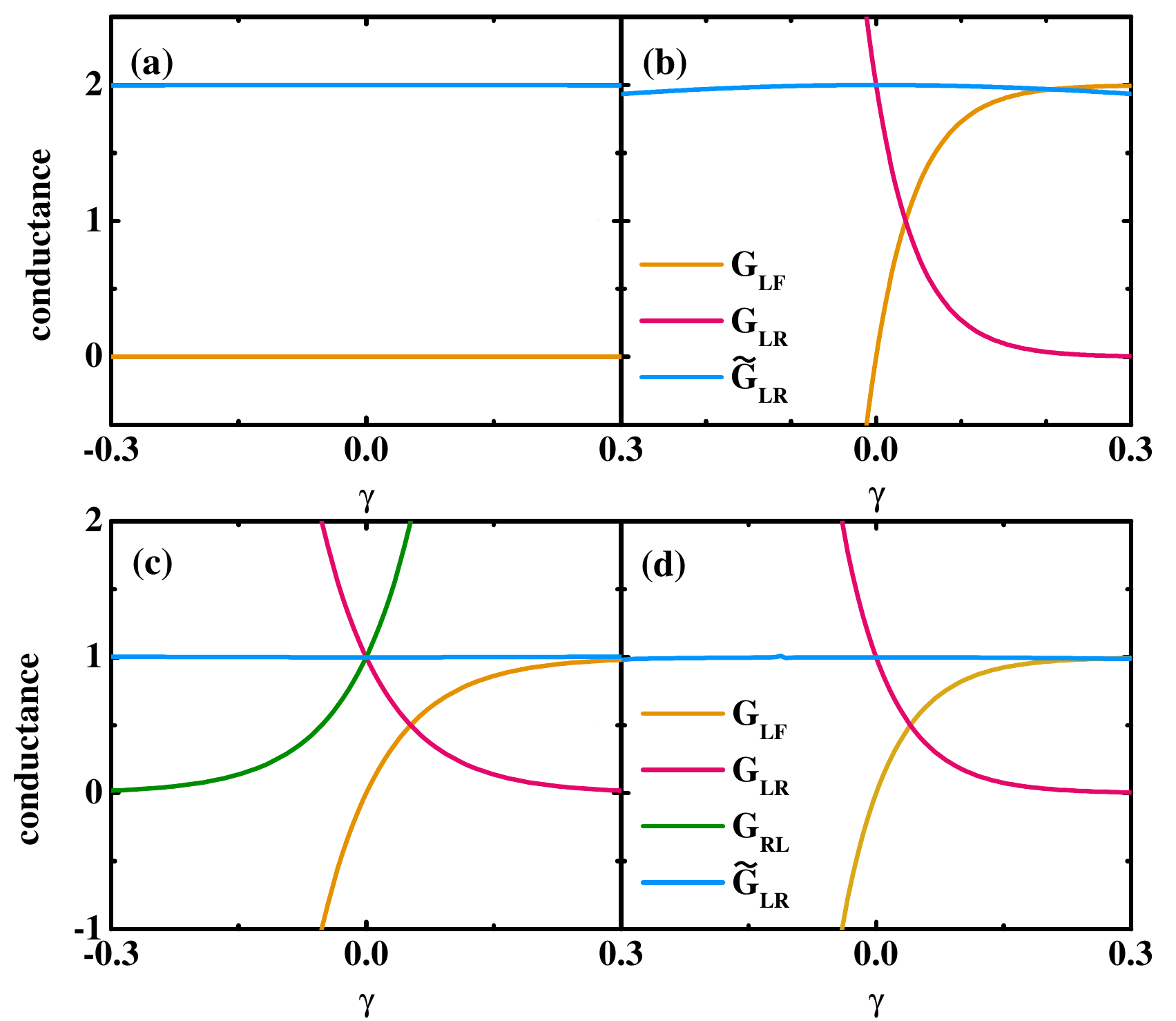}
\centering
\caption{(Color online) Conductances $G_{LR}$, $G_{LF}$ and $\tilde G_{LR}$ as functions of non-Hermitian strength $\gamma$ for different quantum transport systems. (a) BHZ model with complex potential $H_1=i\gamma (\sin k_x)\sigma_x \tau_x$; (b) BHZ model with complex potential $H_1=i\gamma \sigma_0 \tau_0$; (c) Haldane model with complex potential $\gamma_A=-\gamma_B=\gamma$; and  (d) Haldane model with complex potential $\gamma_A=\gamma_B=\gamma$. For each system, the Hamiltonian is non-Hermitian in the central region, while maintains Hermitian in both leads.}
\label{fig2}
\end{figure}	
		
Next, we consider the influence of a complex potential on a topological edge state described by the Bernevig-Hughes-Zhang (BHZ) model\cite{BHZ}, where the Hamiltonian is $\mathcal{PT}$-symmetric and  defined as
\begin{eqnarray}
H_0(\emph{\textbf{k}}) &=& (m + t \cos k_x + t \cos k_y)\tau_z  \nonumber \\
&+& t(\sin k_y) \tau_y + t(\sin k_x) \sigma_z \tau_x.
\label{BHZ}
\end{eqnarray}
Here, $\emph{\textbf{k}}\!=\!(k_x,k_y)$ represents the reciprocal vector, and the Pauli matrices $\sigma_i$ and $\tau_i$ ($i=x,y,z$) represent the spin and orbital degrees of freedom, respectively. $t$ denotes the hopping parameter, and $m$ is the mass parameter. A $\mathcal{PT}$-symmetric complex potential $H_1=i\gamma (\sin k_x)\sigma_x \tau_x$ or a $\mathcal{PT}$-symmetry-broken complex potential $H_1=i\gamma \sigma_0 \tau_0$ is incorporated into $H_0$ in the scattering region, where $\gamma$ represents the strength of the non-Hermitian term. For the former $H_1$, its eigenvalues are real. However, for the later $H_1$, the eigenvalues become complex. Fig.2(a) and 2(b) show the conductances $G_{LR}$, $G_{LF}$ and $\tilde G_{LR}$ for these two different non-Hermitian systems, respectively. In Fig.2(a), $G_{LF}$ is zero regardless of $\gamma$, making $\tilde G_{LR}$ identical to $G_{LR}$. The magnitude of $G_{LR}$ remains at 2, indicating the robustness of the topological edge state to the complex potential. In Fig.2(b), however, $G_{LF}$ is nonzero, leading to a significant difference between $\tilde G_{LR}$ and $G_{LR}$. Firstly, $G_{LR}$ increases as $\gamma$ decreases, even exceeding 2 when $\gamma$ is less than zero. This result is not physically reasonable because there are only two transmission channels in both Hermitian leads. After applying the current partition based on gauge invariance, $\tilde G_{LR}$ remains below 2, which is more consistent with physical expectations. Additionally, $\tilde G_{LR}$ decays slightly as $\gamma$ increases due to the introduction of dephasing, which is consistent with the behavior shown in Fig.1(c). These results illustrate that $G_{LF}$ is zero for a $\mathcal{PT}$-symmetric Hamiltonian but nonzero for a $\mathcal{PT}$-symmetry-broken system.

Finally, we consider a $\mathcal{PT}$-symmetry-broken system described by Haldane model to explore the influence of non-Hermitian to quantum transport. The tight-binding lattice Hamiltonian is given by\cite{Haldane, PRL.128.223903}
\begin{eqnarray}
	H =&& \sum_{\langle i,j\rangle}c_i^\dagger c_j + e^{i\psi}\sum_{\langle\langle i,j\rangle\rangle }c_i^\dagger c_j  +i\gamma_A\sum_{i\in A}c_i^\dagger c_i +  i\gamma_B\sum_{i\in B}c_i^\dagger c_i. \nonumber \\
	\label{Haldane}
\end{eqnarray}
Here, the first term on the right-hand side represents the nearest-neighbor coupling, and the second term denotes the next-nearest-neighbor coupling with a phase factor $\psi$. Complex potentials $i\gamma_A$ and $i\gamma_B$ are incorporated into sites $A$ and $B$, respectively, of the $AB$-sublattice. For this Hamiltonian, its eigenvalues are complex. Analogously, we consider a two-probe system where non-Hermitian terms are included in the central scattering region. Fig.2(c) and 2(d) show the conductance $G_{LR}$, $G_{LF}$ and $\tilde G_{LR}$ as functions of $\gamma$ in two distinct cases: $\gamma_A=-\gamma_B=\gamma$ and $\gamma_A=\gamma_B=\gamma$, respectively. Since $G_{LF}$ is nonzero in both cases, $\tilde G_{LR}$ is different from $G_{LR}$. When $\gamma_A=-\gamma_B$, $G_{LR}$ and $G_{RL}$ exhibit a symmetric distribution around $\gamma = 0$. When $\gamma_A=\gamma_B$, $G_{LR}$ and $G_{RL}$ are identical for various values of $\gamma$. Given that both leads have only one transmission channel, the conductance can not exceed 1. Therefore, it is unreasonable for $G_{\alpha\beta}$ to exceed 1, whereas it is more logical for $\tilde G_{LR}$ to remain below 1.

Our numerical calculations for different non-Hermitian systems verify the validity of the gauge invariant conductance described in Eq.(\ref{cond1}). When the eigenvalues of the non-Hermitian Hamiltonian of the quantum dot $H_0-i\Gamma_F/2$ are real,  $G_{\alpha F}$ vanishes as exemplified in Fig.1(b) and Fig.2(a). In this case, gauge invariance of $G_{LR}$ is naturally satisfied, and the current can be directly calculated using the Laudauer-B$\ddot{u}$ttiker formula. Conversely, if the eigenvalues of the Hamiltonian of the system are complex, as shown in Fig.1(c) and Fig.2(b)-(d), $G_{LR}$ must be replaced by $\tilde G_{LR}$ to maintain gauge invariance.

\section{Summary}

We have derived a current expression for non-Hermitian multi-probe systems using the NEGF method. The current expression is different from the well-established Landauer-B$\ddot{u}$ttiker formula for Hermitian system, featuring an additional term stemming from the non-Hermitian Hamiltonian. As a result, the gauge invariance of conductance may not hold, depending on the value of $G_{\alpha F}$, which signifies the conductance between the real lead $\alpha$ and the virtual probe $F$.
In a pseudo-Hermitian or $\mathcal{PT}$-symmetric non-Hermitian system, $G_{\alpha F}$ vanishes when the eigenvalues of the non-Hermitian Hamiltonian are real. In this case, the current expression reduces to the Landauer-B$\ddot{u}$ttiker formula, naturally ensuring both gauge invariance and current conservation.
However, for a $\mathcal{PT}$-symmetry-broken non-Hermitian system, $G_{\alpha F}$ becomes nonzero, leading to a violation of gauge variance.
To address this, we developed two phenomenological methods: the virtual probe method and the current partition method. Both methods have been successfully implemented to numerically investigate the conductances of various non-Hermitian two-probe systems, thereby ensuring gauge invariance within these systems.

\section{Acknowledgments}
This work was supported by the National Natural Science Foundation of China (Grants No. 12034014) and the Shenzhen Natural Science Foundation (Grants No. 20231120172734001).


\begin{thebibliography}{00}
\bibitem{example1}
For a concrete example of spontaneous emission, the rate of quantum jumps $\gamma$ is roughly $10^9/s$, and the short timescale is much less than 1 ns.
\bibitem{footnote}
In the linear expansion, Eq.(\ref{curr1}) simplifies to $I_\alpha=\sum_\beta G_{\alpha\beta}v_\beta$. When a voltage shift $\Delta V$ is uniformly applied to $v_\alpha$ at each lead, the current changes to $I_\alpha=\sum_\beta G_{\alpha\beta}(v_\beta+\Delta V)$. Due to the gauge invariance, $I_\alpha$ remains unchanged, implying $\sum_\beta G_{\alpha\beta}=0$ since $\Delta V$ is arbitrary.
\end{thebibliography}
\end{document}